\documentclass[11pt,journal,final,a4paper,onecolumn]{IEEEtran}

\usepackage[dvips]{graphicx}
\usepackage{cite}
\usepackage{graphicx}
\usepackage{amsmath}
\usepackage{amssymb}
\usepackage{multirow}
\usepackage{color}
\usepackage{mathtools}
\usepackage{soul}
\usepackage{pdfpages}

\graphicspath{{../Figs/}}
  
\newcommand{\tr}{\rule[-1ex]{0pt}{3.5ex} }

\newcommand{\mk}{$m^*_K$}
\newcommand{\mkss}{$m^*_{K,s-s}$}
\newcommand{\mSx}{$m^*_{\Lambda,x}$}
\newcommand{\mSy}{$m^*_{\Lambda,y}$}
\newcommand{\mSssx}{$m^*_{\Lambda,s-s,x}$}
\newcommand{\mSssy}{$m^*_{\Lambda,s-s,y}$}

\newcommand{\ldaK}{$\lambda_K$}
\newcommand{\ldaKss}{$\lambda_{K,s-s}$}
\newcommand{\ldaS}{$\lambda_{\Lambda}$}
\newcommand{\ldaSss}{$\lambda_{\Lambda,s-s}$}
\newcommand{\IB}{I_{\rm ball}}

\newcommand{\beginsupplement}{%
        \setcounter{table}{0}
        \renewcommand{\thetable}{S\arabic{table}}%
        \setcounter{figure}{0}
        \renewcommand{\thefigure}{S\arabic{figure}}%
     }

\begin{document}
\title{Uniform Benchmarking of Low Voltage Van Der Waals FETs }

\author{Somaia~Sarwat~Sylvia,~\IEEEmembership{Member,~IEEE,} 
        Khairul~Alam,~\IEEEmembership{Member,~IEEE,}
        and~Roger~K.~Lake,~\IEEEmembership{Senior~Member,~IEEE}%
\thanks {S. S. Sylvia and R. K. Lake are with the Department
of Electrical and Computer Engineering, University of California,
Riverside,
CA 92521-0204, USA (e-mail: ssylvia@ece.ucr.edu; rlake@ece.ucr.edu).}
\thanks{K. Alam is with the Department
of Electrical \& Electronic Engineering, East West University, Dhaka,
Bangladesh (e-mail: kalam@ewubd.edu).}
\thanks{
We thank Prof. E. Tutuc for sharing his unpublished
mobility data.
This work is supported in part by FAME, one of six centers of STARnet,
a Semiconductor Research Corporation program sponsored by MARCO and DARPA.
Published in {\em IEEE Journal on Exploratory Solid-State Computational Devices and Circuits}, 
DOI: 10.1109/JXCDC.2016.2619351.
}
}


\maketitle

\begin{abstract}
Monolayer  MoS$_2$, MoSe$_2$, MoTe$_2$, WS$_2$, WSe$_2$, and 
black phosphorous 
field effect transistors (FETs) operating in the low-voltage (LV) regime (0.3V) 
with geometries from the 2019 and 2028 nodes of the 2013 
International Technology Roadmap for Semiconductors (ITRS)
are benchmarked along with an ultra-thin-body Si FET.
Current can increase or decrease with scaling, and the trend is strongly correlated
with the effective mass.
For LV operation at the 2028 node, an effective mass of $\sim0.4$ $m_0$,
corresponding to that of WSe$_2$, gives the maximum drive current.
The short 6 nm gate length combined with 
LV operation is forgiving in its requirements for material quality and contact resistances. 
In this LV regime,
device and circuit performance are competitive using currently measured values for mobilities 
and contact resistances for the monolayer two-dimensional materials.
\end{abstract}


\section{Introduction}\label{sec:intro}
%
%

There is significant interest in understanding how two-dimensional (2D) semiconductors
compare with traditional semiconductors for use as the channel material in ultra-scaled field effect transistors
(FETs). 
The FET also serves as a baseline device for determining targets for material parameters.
For example, given a set of FET performance specifications such as drive current, switching energy, switching
delay, etc., one can then ask, ``What material parameters, such as, for example, mobility, effective mass, bandgap, or
contact resistance, are sufficient to achieve these device performance metrics?''
One can also enquire, ``What material parameters optimize the device performance?''
Thus, benchmarking of a baseline device provides top-down targets for 
materials benchmarking \cite{Kos_IEEE_ESSCDC15}. 

Promising 2D semiconductors include the transition metal dichalcogenides (TMDs) with the chemical
form MX$_2$ where M = Mo or W and X = S, Se or Te \cite{AKis_nano, Falko_allTMDC_bandstructure, MoSe2_mobility_apl_2012, 14_TMDC_mobility, AKis_nnano_review, WSe2_DJena_Nano_Letters, WSe2_pfet_Nano_Letters, MoTe2_FET_APL_2014},
and bandgaps in the range of 1--2 eV \cite{AKis_nnano_review, Falko_allTMDC_bandstructure}.
A more recent addition to the van der Waals (vdW) class of materials 
for field effect transistor (FET) applications is black phosphorus 
(BP) \cite{BP_nature_march_2014, BP_acs_nanao_2014, BP_nat_comm_jul_2014}.
BP's large field effect mobility and highly anisotropic bandstructure 
make it a promising material for FET applications
\cite{Nature_BP_mobility, BP_nature_march_2014, BP_JGuo_EDL_2014, BP_acs_nanao_2014,BP_review_arxiv, renaissance_BP_PNAS}.  

A number of articles in the literature have theoretically predicted the performance of 
these alternate materials for future device applications.
While the majority of the performance predictions are for MoS$_2$ FETs 
\cite{Sayeef_MoS2_how_good, JGuo_MoS2_scaling_lmt, kalam2012mos2, MoS2_FET_strain_JAP_2013, Luisier_MoS2_PRB_eph_scatt, Banerjee_MoS2_APL_2013, MoS2_strained_TED_2013, sayeef_mos2_full_band_2015} and BP \cite{BP_JGuo_EDL_2014}, 
some of them focus on device comparisons  within the TMD group for conventional FETs 
\cite{TMD_FET_2014benchmarking,JGuo_TMD_FET_TED_2011, Banerjee_JAP_TMD_FET} 
and for tunnel FETs \cite{Klimeck_TMD_TFET,JGuo_TMD_TFET}. 
The BP FET was compared against the MoS$_2$ FET in Ref. \cite{HGuo_BP_vs_MoS2_TED_2014}.
A BP based TFET was proposed in Ref. \cite{klimeck_bp_tfet}.

There are two different operation regimes denoted as high performance (HP) and low power (LP) defined in the 
2013 ITRS \cite{ITRS_13}.
There is also a low voltage (LV) regime considered in Ref. \cite{Young_LV11}
and benchmarked in Refs. \cite{young_nikonov_bcb2, young_nikonov_bcb3}. 
It is this LV regime that we consider in this work with a supply voltage of 0.3 V. 
As of  today, there are a large number of material candidates for future CMOS devices.
But little is known about their relative performance in the LV regime, since, to the best of our knowledge, 
they have never been compared in a single systematic study.
In general,  LV has been given less attention than HP or LP operation.

Inspired by the device benchmarking of the Nanoelectronics Research Initiative 
(NRI) \cite{young_nikonov_bcb2, young_nikonov_bcb3} 
and the materials benchmarking of STARnet centers \cite{Kos_IEEE_ESSCDC15},
in this work we present and compare BP and 5 different TMD based FETs.
For a baseline comparison, we also simulate an ultra thin body (UTB) Si FET using the same 
model and code. 
The vdW materials that we chose to compare are MoS$_2$, MoSe$_2$, MoTe$_2$, WS$_2$, WSe$_2$ and BP.
%
%
Performance metrics are compared for individual devices as well as for a 
standard integrated circuit of a 32 bit adder.  
using the beyond CMOS benchmarking (BCB) scheme 3.0 \cite{young_nikonov_bcb3}.

\section{Simulation Method}
\label{sec:computation}
The structural parameters for the devices were taken from columns 2019 and 2028 of the 
Low Power (LP) technology requirement tables, 
ITRS 2013 \cite{ITRS_13}. 
The values are summarized in Table \ref{tab:dim}.
The devices are assumed low voltage (LV) with $V_{\rm DD}$ = 0.3 V \cite{young_nikonov_bcb3}. 
Two different production years were selected to examine the 
effect of scaling on the devices of interest. 
We primarily considered single gate (SG) FETs, 
and a few exemplary simulations were performed for double gate (DG) structures as well.

\begin{table}
\caption{Device dimensions according to ITRS 2013 \cite{ITRS_13}}
\vspace{-18pt}
\label{tab:dim}
\begin{center}
\begin{tabular}{|c|c|c|}
\hline
\tr  Structural Parameters                                             & \multicolumn{2}{|c|}{Year of production} \\
\cline{2-3}
\tr                                                                                & 2019 & 2028\\
\hline
\tr Metal 1 1/2 pitch, F (nm)                                         & 20     & 7.1 \\
\hline
\tr Physical gate length, $L_g$ (nm)                             & 13.3  & 5.9 \\
\hline
\tr Effective channel length, $L_{ch}$ (nm)                   & 10.6 & 4.7\\
\hline
\tr Dielectric constant of top gate oxide, $\epsilon_r$  & 15.5 & 20\\
\hline
\tr Physical gate oxide thickness, $t_{ox}$ (nm)           & 2.42 & 2.10\\
\hline
\tr Equivalent oxide thickness, $EOT$ (nm)                  & 0.6089 & 0.4095\\
\hline                
\end{tabular}
\end{center}
\end{table} 

\begin{figure}
\centering
\includegraphics[width = 2.5in]{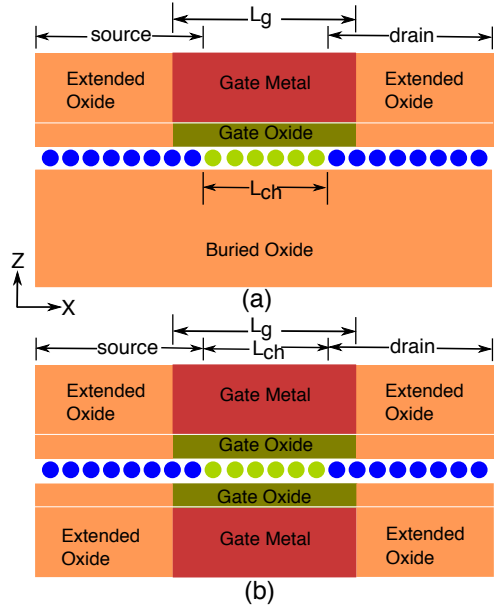}
\caption{Cross section of the device used for simulation (a) SG FET (b) DG FET. The central line of dots indicate the monolayer device region.}
\label{fig:dev}
\end{figure}

Figure \ref{fig:dev} shows the device structures used for the simulations in this work \cite{Klimeck_xsistor_roadmap4_ITRS_2013}.
The buried oxide and extended oxide regions are SiO$_2$ with a dielectric constant of 3.9.
The gate oxide is composed of both high-K (according to Table \ref{tab:dim}) oxide under the gate 
and SiO$_2$ \cite{Luisier_IEDM2011} in the source-drain extensions for improved gate control.
For the Si FET, transport from source to drain is in the (100) direction.
For the BP FET, transport is in the $X$ direction, the direction of the light mass.
For the circuit metrics, the default width of 4 times the pitch is used for the 
FETs \cite{young_nikonov_bcb3}. 

For the TMD and Si FETs, electron conduction is considered while for 
the BP FET, 
{
both electron and
}
hole conduction are considered, 
since most recent experimental work focuses on hole transport \cite{BP_review_MatChemC}.
For the vdW materials, 
the source and drain doping densities were swept from $1\times10^{19}$ to
$1\times10^{20}$ cm$^{-3}$  ($\sim5.7 \times 10^{11} - \sim 7.3 \times 10^{12}$ cm$^{-2}$). 
For each node and geometry, two results are recorded.
One result is for the doping density that results in the maximum
drive current.
The second result is for the maximum doping density of $1\times10^{20}$ cm$^{-3}$.
The drive currents versus source doping are shown in Fig. S1
of the Supplementary Information.
{
This optimization is performed with the contact resistance set to zero.
}
For the 3 nm Si UTB FETs, a source and drain doping density of
$1\times 10^{19}$ cm$^{-3}$ ($3 \times 10^{12}$ cm$^{-2}$) is used \cite{kalam2012mos2}.

Material properties for all of the materials considered in this work are 
summarized in Table S1 of the Supplementary Information.
The UTB Si has a finite thickness of 3 nm, and  
all of the vdW materials are monolayers. 
It has been shown that adding multiple layers on top of a single layer 
cannot boost the on current \cite{sayeef_mos2_full_band_2015}.
Listed mobilities for monolayer vdW materials are experimentally measured values  
obtained from the literature \cite{MoS2_1L_natcomm, MoSe2_monolayer_acsnano, WS2_1L_BNsandwich, WSe2_DJena_Nano_Letters, BP_1L_sdas_nanolett,BP_1L_extracted_mu_nat_comm} except for MoTe$_2$.
Mobility in monolayer MoTe$_2$ was unknown at the time of this work, 
hence it was approximated from MoSe$_2$ using both materials' electron 
effective masses (see footnote of table S1). 

\begin{figure}
\centering
\includegraphics[width = 3in]{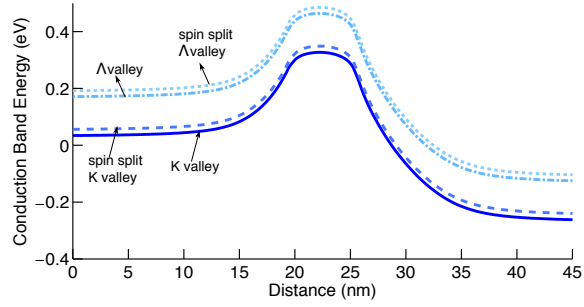}
\caption{Illustration of the multiple single-band approach used in this work. Potential profile for 4 valleys in the conduction band for a typical TMD FET is shown. }
\label{fig:mult_single_band}
\end{figure}

As evident from the conduction band $\Lambda$-valley to $K$-valley energy separation, 
$\Delta_{K\Lambda}$,
listed in Table S1, 
all values of $\Delta_{K\Lambda}$ are less than $V_{DD}$, and, therefore they
will have an effect on the electron transport in the TMD FETs.
Also, there is considerable spin-splitting in many of the conduction band $K$ and $\Lambda$ valleys.
Therefore, we have taken the different spins and valleys into account by using a 
multiple single-band approach as depicted in Figure  \ref{fig:mult_single_band}.
In this approach, each spin and valley is treated as an independent band with its own
effective mass.

For each band, the discretized effective mass Schr\"{o}dinger equation 
is solved for the charge density using 
a non-equilibrium Green function (NEGF) approach
similar to that described in \cite{kalam2012mos2}.
The heavily doped source and drain regions are treated as contacts in equilibrium with
their respective Fermi levels \cite{Klimeck_APL95}.
The total charge at each site is the sum of the charge calculated for each band.
The charge is self-consistently solved with Poisson's equation.
The electrostatic potential within the device is calculated using a 
2D finite difference solution of Poisson's equation discretized on a 
0.2 nm grid within the channel and a 0.5 nm grid within the oxide.
Dirichlet boundary conditions are set at the metal gate and Von Neumann boundary 
conditions are used at all other exterior boundaries.

Once the charge calculation has converged, 
current is calculated for each band. 
The contribution from all bands is summed to give the total current.
The effect of scattering in the channel is included with a reflection coefficient 
determined from a mean free path related to the mobility and an effective channel length
\cite{lundstrom_elem_backscattering_th,lundstrom_backscattering}.
Details are provided in the Supplementary Information.

The off current is set at  1.5 $\rm nA/\mu m$ for all devices.
The drain bias $V_{DS}$, and on-state gate voltage $V_{GS}$ are 0.3 V.
The maximum allowable source-drain total contact resistances ($R_{SD}$) 
are estimated following the methodology used by 
the ITRS \cite{ITRS_13, Klimeck_xsistor_roadmap4_ITRS_2013}.
For this, a reference value of current was first calculated with scattering included but $R_{SD}$ set to 0.
A set of simulations including scattering were then performed for a range of 
$R_{SD}$ values.
$R_{SD}$ was divided equally between the source and drain.
In the self-consistent loop, the internal gate and drain potentials with respect to the source,
$V_{GS}^\prime$ and $V_{DS}^\prime$,
were updated at each iteration 
according to  $V_{GS}^\prime = V_{G} -  I_D R_{SD} / 2$ and 
$V_{DS}^\prime = V_{DD} - I_D R_{SD}$, where $V_G$ is the applied gate voltage
with respect to ground.
The series resistance raises the source potential by $I_D R_{SD} / 2$ which lowers
the gate to source voltage by the same amount.
The particular value of $R_{SD}$ that resulted in a 33.3\% reduction of current compared to the reference current was then chosen as the maximum allowable contact resistance for the LV devices.

Two performance metrics are the switching delay and the switching energy
defined as
\begin{equation}
t = CV_{DD}/I_{\rm on}~  \text{and~} E = CV_{DD}^{2} .
\label{eq:E&t}
\end{equation}
Here, $I_{\rm on}$ is the on-current, 
$C$ is the total capacitance that includes the oxide capacitance, 
the semiconductor capacitance (also known as quantum capacitance), 
and any parasitic capacitance that might be present.
The capacitance is determined as follows.
The total capacitance $C = \partial Q/\partial V_{G}$ where $Q$ is the total charge in the entire semiconductor region that includes the source, channel and drain.
In this manner, the gate and fringing capacitances are taken into account all at the same time.
In doing so, one has to make sure that no other external inputs are changing except the applied gate bias.
Therefore, the total charge $Q$ is calculated with
$R_{SD}$ = 0, since
$R_{SD}$ alters the effective gate voltage $V_{GS}^\prime$.  

{
The calculated drive currents and capacitances 
are input into the BCB 3.0 scripts} \cite{young_nikonov_bcb3}.
{
The BCB 3.0 scripts use the input for one type of transistor
and approximate the on-current of the 
pFET is equal to that of the nFET.
Delay times and switching energies are calculated using empirical rules
chosen to match SPICE simulations. 
For circuits, a per unit length interconnect capacitance of 126 aF/$\mu$m is used, and the
interconnect length associated with each transistor is $20F$ where $F$
is the DRAM half pitch corresponding to the technology node.
Full details of the BCB 3.0 method are given in Ref.} \cite{young_nikonov_bcb3}.

\section{Results}
\label{sec:results}

\begin{table*}
\caption{Ballistic on currents, $I_{\rm ball}$ ($\rm \mu A/\mu m$) and scattering limited on currents, 
$I_{\rm scatt}$ ($\rm \mu A/\mu m$) for both 2019 and 2028 nodes. 
Source drain doping, $N_{SD} \rm (cm^{-3})$ is the optimum doping at which $I_{\rm scatt}$ maximizes for the vdW FETs.
For $\rm Si$, current maximizes at a doping even lower than $1 \times 10^{19} \rm cm^{-3}$ and hence this value was chosen as a compromise between current and screening length. 
$\IB$ was calculated at the listed $N_{SD}$'s where contact resistance, $R_{SD}$ ($k\Omega\mu m$) and backscattering coefficient, $r_c$ were both set to 0. 
$I_{\rm scatt}$ is the on current where both $R_{SD}$ and $r_c$ are included. BP(N) and BP(P) refer to n-type and p-type BP FETs, respectively.
}
\vspace{-10pt}
\label{tab:Ion}
\begin{center}
\begin{tabular}{c c|c|c|c|c|c|c|c|c|}
\cline{3-10}
\tr & & \multicolumn{4}{|c|}{13.3 nm} & \multicolumn{4}{|c|}{5.9 nm} \\
\cline{3-10}
\tr  &                                           & $I_{\rm ball}$ & $I_{\rm scatt}$ & $N_{SD} (\times 10^{19})$ & $R_{SD}$ & $I_{\rm ball}$ & $I_{\rm scatt}$
                                                                                                                   & $N_{SD} (\times 10^{19})$ & $R_{\rm SD}$ \\

\hline
\multicolumn{1}{|c|}{MoS$_2$} &
\multicolumn{1}{|c|}{SG}              &\tr 63.14    & 20.35      & 3      & 1.68      & 56.86    & 25.21   &  3    & 1.2\\
\cline{2-10}
\multicolumn{1}{|c|}{} &
 \multicolumn{1}{|c|}{DG}           &\tr 107.1    & 33.6        & 6       & 0.9        & 110.2     & 48.06  & 5     & 0.6\\
\hline
\multicolumn{2}{|c|}{MoSe$_2$} &\tr 61.81    & 17.89      & 4       &1.8         & 56.8      & 22.75   & 3     & 1.28\\
\hline
\multicolumn{2}{|c|}{MoTe$_2$}&\tr 61.7      & 16.9        & 4       & 1.9         & 55         & 21.16  & 3      & 1.57\\
\hline
\multicolumn{1}{|c|}{WS$_2$} &
\multicolumn{1}{|c|}{SG}           &\tr 64.6       &27.18       & 3       &1.1          & 62         & 30.44  & 2      & 1.03\\
\cline{2-10}
\multicolumn{1}{|c|}{} &
 \multicolumn{1}{|c|}{DG}         &\tr 109.8     & 45.18      & 4       & 0.63       & 113.2     & 56.9   & 4       & 0.5\\
\hline
\multicolumn{1}{|c|}{WSe$_2$}&
\multicolumn{1}{|c|}{SG}           &\tr 67.6       & 29.67      & 3       &1.02        & 64.85     & 34.2   & 2       & 0.83\\
\cline{2-10}
\multicolumn{1}{|c|}{} &
 \multicolumn{1}{|c|}{DG}          &\tr 116     & 50.23       & 4           & 0.53        & 116.56       & 62.5  & 5  & 0.43\\
\hline
\multicolumn{1}{|c|}{BP(N)}&
\multicolumn{1}{|c|}{SG}           &\tr 70.65     & 26.14      & 3       & 1.18      & 58.94          & 26.91 & 2.1  & 1.11\\
\cline{2-10}
\multicolumn{1}{|c|}{} &
 \multicolumn{1}{|c|}{DG}        &\tr 121.6      & 42.08      & 4       & 0.76     & 111.15         & 50.95 & 4  & 0.6\\
 \hline
\multicolumn{1}{|c|}{BP(P)}&
\multicolumn{1}{|c|}{SG}           &\tr 75.8     & 33.45      & 2.55       &0.95      & 59.2          & 30.55 & 2  & 0.95\\
\cline{2-10}
\multicolumn{1}{|c|}{} &
 \multicolumn{1}{|c|}{DG}        &\tr 123.4      & 53.26      & 4       & 0.54     & 113.5         & 58.4 & 4  & 0.53\\
\hline
\multicolumn{1}{|c|}{Si} &
\multicolumn{1}{|c|}{SG}           &\tr 19.7    &9.73           & 1        &2.9        & 6.22           & 3.73 & 1   &6.5\\
\cline{2-10}
\multicolumn{1}{|c|}{} &
\multicolumn{1}{|c|}{DG}           &\tr 77.25 &36              & 1        &0.87      & 45.46        & 25.07 & 1  &1.1\\
\hline
\end{tabular}
\end{center}
\end{table*}

The on-current, optimum doping, and series resistance for each material,
node, and geometry are tabulated
in Table \ref{tab:Ion}.
$I_{\rm ball}$ refers to the 
ballistic on-current calculated with both the contact resistance 
$R_{\rm SD}$ and the backscattering coefficient $r_c$ set to 0.
$R_{SD}$ is the maximum allowable total contact resistance 
(source plus drain)
that degrades the current calculated in the presence of scattering
by 33.3\%. 
$I_{\rm scatt}$ is the on-current where both $r_c$ and the maximum allowable 
$R_{SD}$ are included. 
For the rest of our discussion, unless otherwise noted, 
the on-current will refer to the scattering limited current, $I_{\rm scatt}$.
\begin{figure}
\centering
\includegraphics[width = 3.2in]{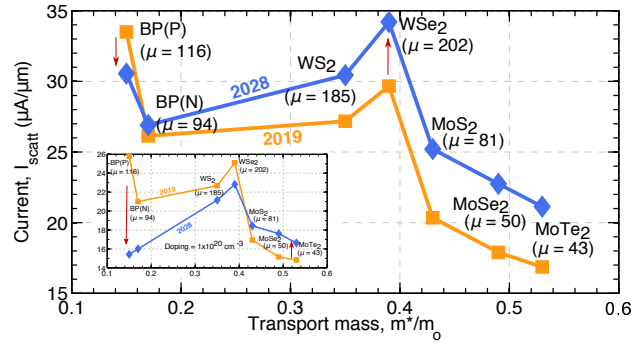}
\caption{
$I_{\rm scatt}$ versus effective mass for the SG vdW FETs at the 2019 and 2028 nodes
with optimized source and drain doping.
Each data point is labeled with its material and mobility.
The red arrows indicate the current 
trends when scaling the gate lengths from the 2019 to the 2028 node.
Inset shows the same plot for a fixed doping density of $1 \times 10^{20}~\rm cm^{-3}$.
}
\label{fig:I_vs_m}
\end{figure} 
For the SG vdW materials, the drive currents at optimum doping are shown in 
Fig. \ref{fig:I_vs_m} plotted versus the bandedge effective mass of the material
for both the 2019 and the 2028 nodes.
The inset is the same plot but for a fixed doping of $1.0 \times 10^{20}$ cm$^{-3}$.
Each data point is labeled with its material.
%

%
The physical mechanisms governing FET performance are the same as those 
analyzed in Ref. \cite{Young_JEDS14} for III-V FETs, the balance between source
exhaustion and tunneling leakage.
The range of transport effective masses, from $0.15~ m_0$ for 
{p-type}
BP to $0.53~ m_0$
for MoTe$_2$, make this balance different for the different materials.
The optimum source doping is lower for the lighter mass materials.
The lower doping results in longer screening lengths of the channel potential
into the source and the drain regions increasing the effective channel length
and decreasing the off-state direct tunneling.

\begin{figure}
\centering
\includegraphics[width = 2.8in]{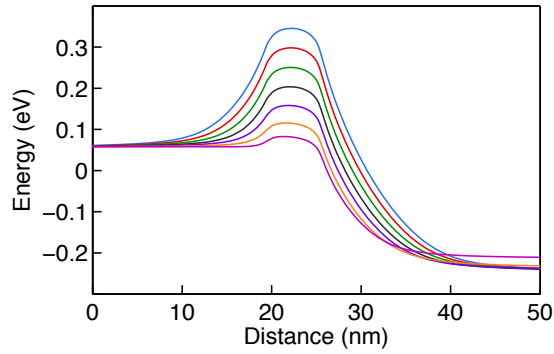}
\caption{
Energy band diagrams for different gate biases of the 
{p-type}
BP FET at the 2028 node with doping density of $2\times 10^{19}$ cm$^{-3}$.
{Kinetic energy for the holes is taken to be positive.}
The source Fermi energy is 0 eV.
}
\label{fig:BP_Ec_2e19}
\end{figure}

For the X-directed transport in 
{p-type}
BP, the low mass in the
transport direction provides a high velocity, and the
large transverse effective mass provides many modes 
for transport. 
%
Because of the low transport mass, the optimum source doping of
$2\times 10^{19}$ cm$^{-3}$ is the lowest among the vdW FETs.
As shown in Fig. \ref{fig:BP_Ec_2e19},
in the off-state, the low doping results in long screening lengths of the channel potential
into the source and the drain regions 
increasing the effective off-state channel length
and decreasing the off-state direct tunneling.
The off-state channel potential decays approximately 10 nm into the
source and 15 nm into the drain giving an off-state total
effective source to drain length of 30 nm at the 2028 node.
In the on-state, the small channel potential decays within a few nanometers
into the source, and the high field region extends approximately 10
nm into the drain.
Thus, the effective source to drain region in the on-state is 
approximately 15 nm.
One advantage is that the longer depletion lengths in the source and drain reduce the
fringing capacitance between the source and drain and therefore
reduce the RC delay time.
A disadvantage is that the transit time increases.
A saturation velocity of $10^7$ cm/s gives a
transit time that is
10 times less than the RC delay time. 
At $10^6$ cm/s, the two times are comparable.

As the gate length is scaled from 13 nm to 6 nm, 
with optimized doping, 
the on-current of BP drops slightly,
and the on-currents of all of the TMD FETs increase.
The TMD FETs with the heavier effective masses benefit from scaling,
while the BP FET with the lightest transport mass is degraded by the scaling.
%
%
In every case, the ballistic current decreases as the channel length
decreases from 13 nm to 6 nm, in agreement with previous 
work \cite{Klimeck_xsistor_roadmap4_ITRS_2013},
and the ballistic current of 
{p-type}
BP with the lightest transport mass
decreases the most.
For BP, the large decrease in the ballistic current dominates, and the
total current including scattering decreases.
For the heavier mass TMDs, the ballistic current is only
slightly reduced.
As the channel length becomes comparable to the mean
free path, reflection is reduced. 
This process dominates for the TMDs with heavier effective masses, 
and their on-current increases
as the gate length is scaled down to 6 nm.

\begin{figure}
\centering
\includegraphics[width = 3.2in]{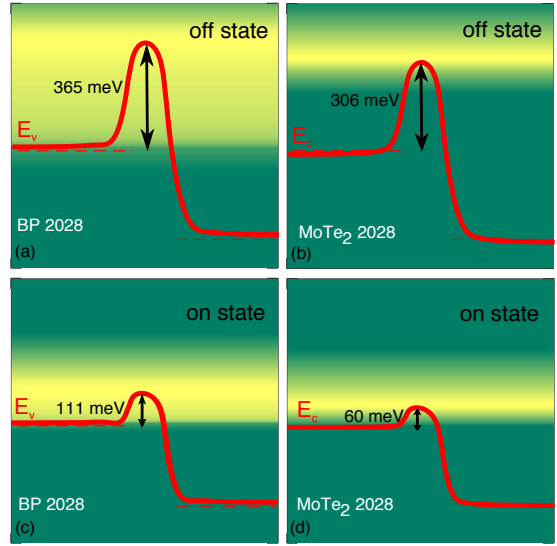}
\caption{(Color online)
(a) 
{
Valence band edge for p-type
} 
BP 
{(with the hole energy taken as positive)} 
and (b) 
the conduction band edge of MoTe$_2$ 
{(with the electron energy taken as positive)}
in the off-state and 
(c) BP and (d)  MoTe$_2$ in the on-state for the 2028 node 
with fixed source and drain doping of $1.0\times 10^{20}$ cm$^{-3}$.
The source Fermi energy is the reference energy at $E=0$.
The background color indicates the current density per unit energy on 
a log scale. 
Yellow is the highest current.
}
\label{fig:BP_MoTe2_Ec}
\end{figure}

The effective mass affects two processes that determine if the current
will increase or decrease with scaling, and the trends become very clear
with a fixed source and drain doping of $1 \times 10^{20}$ cm$^{-3}$
as shown in the inset of Fig. \ref{fig:I_vs_m}. 
The first process is direct tunneling through the channel, 
and the second process is scattering in the channel.
The process of direct tunneling is governed by the effective mass of the channel material.
A heavier mass minimizes the off-state leakage which enhances the drive current for a
fixed $V_{DD}$, because a smaller percentage of $V_{DD}$ is required to shut the device off. 
This effect is illustrated in Fig. \ref{fig:BP_MoTe2_Ec}. 
The background color indicates the current spectrum (on a log scale) 
with the brightest yellow indicating the highest current.
A comparison of Figs. \ref{fig:BP_MoTe2_Ec}(a) and (b) shows that, in the off-state,
tunneling is significant through
the BP barrier but is suppressed in the MoTe$_2$ barrier.
For BP with the lightest transport mass, the barrier height required to attain the off-state
current of 1.5 nA/$\mu$m is 365 meV.
For MoTe$_2$ with the heaviest transport mass, the barrier height required to attain the 
same off-state
current is 307 meV, approximately 60 meV lower than that for BP.
Applying 0.3 V to the gate reduces the potential in the channels by 254 meV for BP and
247 meV for  MoTe$_2$, so that the barrier height in the on-state is 111 meV for BP
and 60 meV for MoTe$_2$.
Thus, the barrier height of the channel in the on-state for BP is almost twice that
for MoTe$_2$. 
This effect is responsible for the reduction in $I_{\rm ball}$ as the gate length is
scaled from 13 nm to 6 nm. 

The second process of scattering in the channel is also strongly correlated with the
effective mass.
A heavy mass is associated with a short mean free path, so that as the channel
is scaled down to 6 nm, the device becomes more ballistic, $r_c$ decreases, 
and the current increases with scaling.
The Mo compounds have the highest effective masses, the lowest measured electron mobilities,
and the shortest mean free paths as shown in Table S1.
Therefore, these materials benefit most from scaling, 
since direct leakage through the channel
is not a problem, and they become more ballistic as the channel length is scaled.
For BP with the lightest mass in the transport direction, 
the first process of tunneling dominates the performance, 
and there is significant reduction in $I_{\rm scatt}$ going from the 2019 to the 2028 node
when the doping is fixed at $1.0\times 10^{20}$ cm$^{-3}$.
Even at the optimum doping condition, BP is the only 2D material that 
suffers from a reduction in current after scaling.

Adding a second gate to create a DG structure increases the magnitude of the
current, 
and the increase in the magnitude of the current is qualitatively
different for the vdW channels and the UTB Si channel.
At the 2019 node, adding a second gate 
increases $I_{\rm ball}$ by a factor of 1.7 
for the TMD FETs 
and 1.63 for 
{p-type}
BP. 
The increase in $I_{\rm scatt}$ is slightly less.
For 2028 TMDs, adding the second gate increases 
$I_{\rm ball}$ by factors of 1.8 - 1.94 for TMDs and  
{1.9 for both}
BP. 
The increase in $I_{\rm scatt}$ is identical to the increase in $I_{\rm ball}$
within numerical error.
The larger increases in current due to doubling the gates in the 2028
2D FETs indicate that the single gate is losing some control of the
channel when the gate is scaled down to 5.9 nm.
In the DG geometry, the second gate provides greater electrostatic control of the channel. 
The increased gate control 
moves the position where $\Delta V_{\rm ch} = k_BT/q$ further towards
the drain which increases $L_{\rm eff}$ and, consequently, $r_c$,
and is the reason why the increase in $I_{\rm scatt}$ resulting from a second
gate may not be quite as large as the increase in $I_{\rm ball}$.

The maximum allowable projected total contact resistance (source plus drain) $R_{SD}$ for each node 
and material are also included in Table \ref{tab:Ion}.
For the SG devices, 
the current is small, and one can get away with relatively high contact resistances
on the order of 
0.48 to 0.95 k$\Omega \mu$m per contact at the 2019 node, and
0.42 to 0.52 k$\Omega \mu$m per contact at the 2028 node.
To achieve the higher current densities of the DG TMD devices, 
lower contact resistances are required, 
on the order of 
265 - 450 $\Omega \mu$m per contact at 2019 node and
215 - 300 $\Omega \mu$m per contact at 2028 node.
Contact resistances of 240 $\Omega \mu$m 
have already been reported in literature \cite{MoS2_nmat_phase_engg_contact}.

\begin{figure}
\centering
\includegraphics[width = 3.2in]{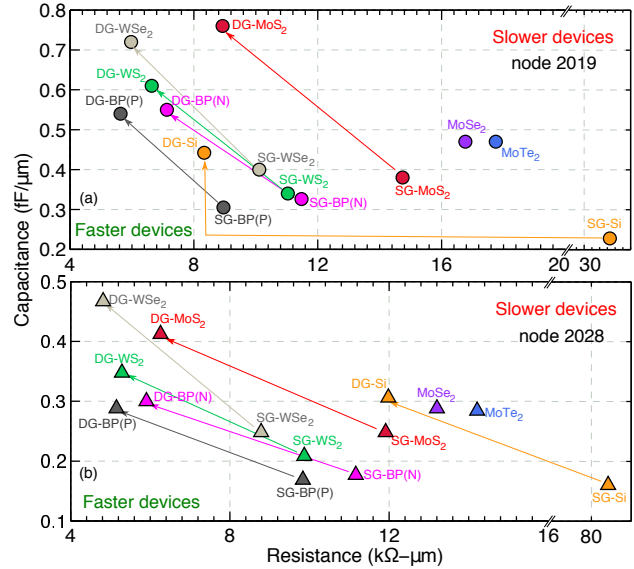}
\caption{Capacitance versus on resistance for individual FETs (a) node 2019 (b) node 2028. 
Data points marked with DG represent double gate structures. 
Circles represent the 2019 node and triangles represent the 2028 node. 
Arrows show the effect of adding a second gate.
}
\label{fig:CvsR}
\end{figure}

From Eq. (\ref{eq:E&t}),
the product of device capacitance and resistance gives the switching delay 
of each individual device.
Fig. \ref{fig:CvsR} shows the capacitance versus resistance for each material,
node, and geometry.
The arrows show the effect of going from a SG geometry to a DG geometry.
First, we discuss the SG geometry at each node.
At the 2019 node, among the SG vdW FETs, 
MoSe$_2$ and MoTe$_2$ have both the most resistance and capacitance and BP has the least.
At the 2028 node, among the SG vdW FETs,  
WSe$_2$ has the smallest resistance among all the vdW materials
since it has the highest drive current, and
BP has the lowest capacitance. 
To understand the low capacitance,
recall that the `device' capacitance is determined by $C = \partial Q / \partial V_G$.
Therefore, if the device is only weakly turned on, there is little charge in the
channel, and $C$ is small, irrespective of the actual geometrical gate 
capacitance. 
Considering the band diagram of BP at the 2028 node in Fig. \ref{fig:BP_Ec_2e19}, 
it is weakly turned on since the top of the barrier is 83 meV above the source Fermi level.
In comparison, MoTe$_2$ with the heaviest mass is more strongly turned on, 
and its capacitance is the highest even though its current is the lowest
among the vdW FETs.
Its low current or high resistance result from the low mobility and short 
mean free paths. 

Both the 2028 SG and DG Si FETs stand out in Fig. \ref{fig:CvsR}.
Applying a DG to 2028 UTB Si gives a capacitance  that is slightly
below the DG vdW FET using 
{p-type}
BP.
There are several reasons for the low capacitance of the Si DG FET.
The 3 nm thick channel requires a double gate to accumulate significant
charge in the channel and turn the device on.
Even when charge is accumulated in the channel, 
the relatively lower effective mass of the lowest quantized state
in the channel of 0.22 $m_0$ results in a lower 
quantum capacitance \cite{kalam2012mos2}.
Finally, the lower doping of the source and drain of $10^{19}$ cm$^{-3}$
compared to the doping of the DG vdW FETs of $4 \times 10^{19}$ cm$^{-3}$ -  
$5 \times 10^{19}$ cm$^{-3}$ results
in longer depletion regions in the source and drain that reduce the 
fringing capacitance for the sidewalls of the gates.
The UTB Si band diagrams shown in Fig. S3 
illustrate these points.
\begin{figure}
\centering
\includegraphics[width = 3.2in]{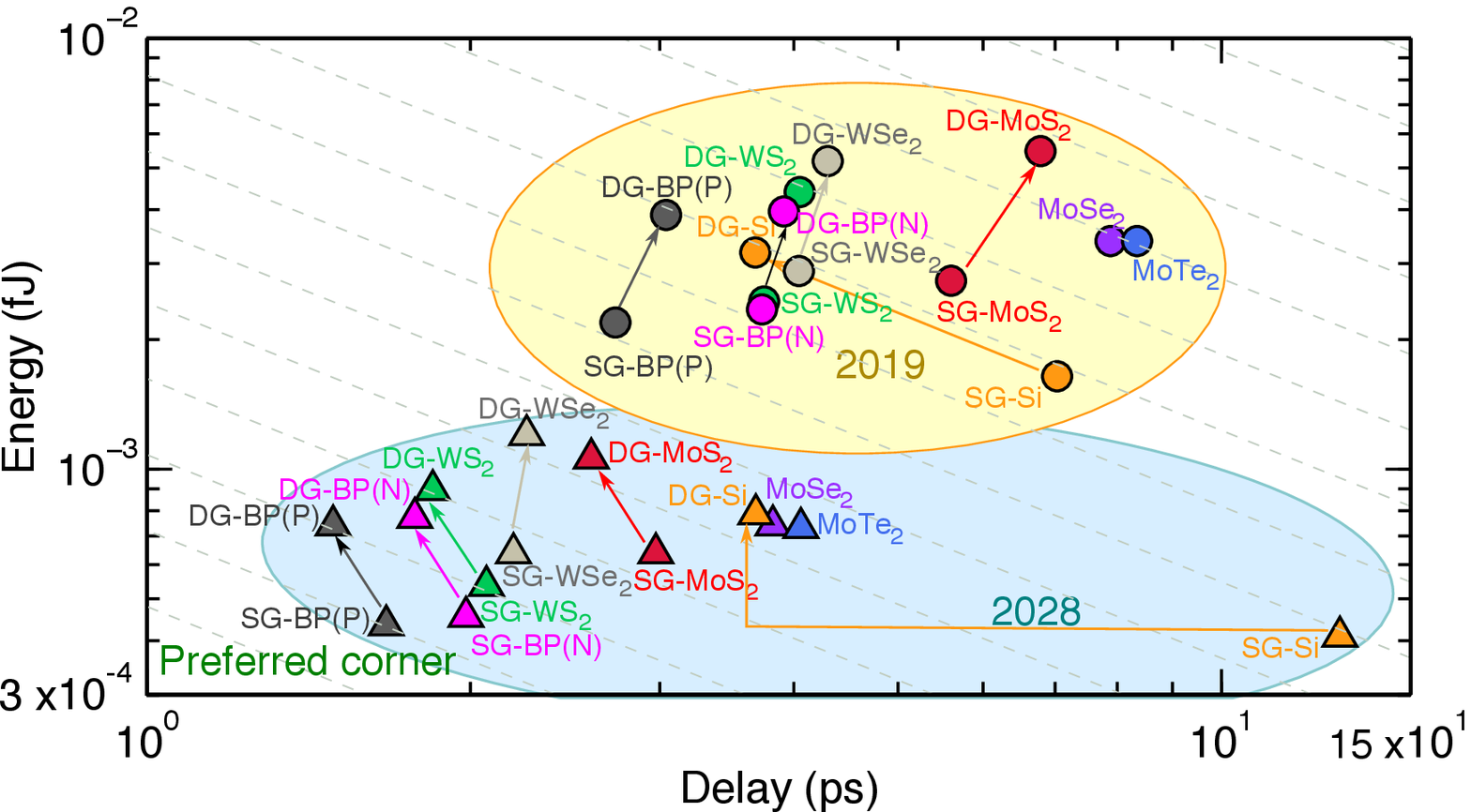}
\caption{Intrinsic switching energy versus delay for individual FETs. 
Circles and triangles stand for the 2019 and 2028 nodes, respectively. 
Diagonal dashed lines are constant energy-delay product lines.
Each successive line represents an increase of 1.5.
\cite{note:bigfig}
}
\label{fig:devEvsT}
\end{figure} 

The intrinsic switching energies versus switching delay times
are shown in Fig. \ref{fig:devEvsT}.
At node 2019, the SG WS$_2$ and WSe$_2$ FETs and 
DG-Si have very similar switching energies and delay times.
Adding a second gate to the 2D materials is detrimental in all cases causing both
the energy and delay to increase.
At the 2028 node, adding a second gate still moves all of the 2D materials to a
higher energy-delay product. 
Only Si is moved to a lower energy-delay product by the addition of a second gate.

Energy-delay benchmarks for a 32 bit adder are shown in Fig. \ref{fig:EvsT}.
Now, the added capacitance of the interconnects is included.
For a per unit length capacitance of 126 aF/$\mu$m, 
the interconnect capacitance per transistor $(c_i)$ is $50$ aF at the 2019 node and 
18 aF at the 2028 node.
The default widths used for the FETs are 4 times the pitches, 
and they are 80 nm at the 2019 node and 28.4 nm at the 2028 node. 
Multiplying these widths times the capacitance values in Fig. \ref{fig:CvsR} gives the 
actual FET capacitances. 
For the vdW FETs,
at the 2019 node, $c_i$ ranges between 1.33 - 2.05 times the SG-FET capacitances and 
between 0.82 - 1.16 times the DG-FET capacitances. 
At the 2028 node, $c_i$ ranges between 2.18 - 3.73 times the SG-FET capacitances and 
between 1.35 - 2.18 times the DG-FET capacitances. 
The interconnect contribution to the delay 
depends on the current that flows through the interconnect, 
and this current is the same as the device current.
As a result, the drive current becomes more important 
for the performance of circuits.
For a SG-TMD FET at either the 2019 or 2028 node, 
adding a second gate increases the intrinsic device switching energy
more than it decreases the delay, so that the device energy-delay product increases.
This same trend applies to the 2019 circuit.
However, for the 2028 circuit, 
adding a second gate leaves the energy-delay product almost
unchanged for BP, WS$_2$ and MoS$_2$ and slightly increased for WSe$_2$.

\begin{figure}
\centering
\includegraphics[width = 3.2in]{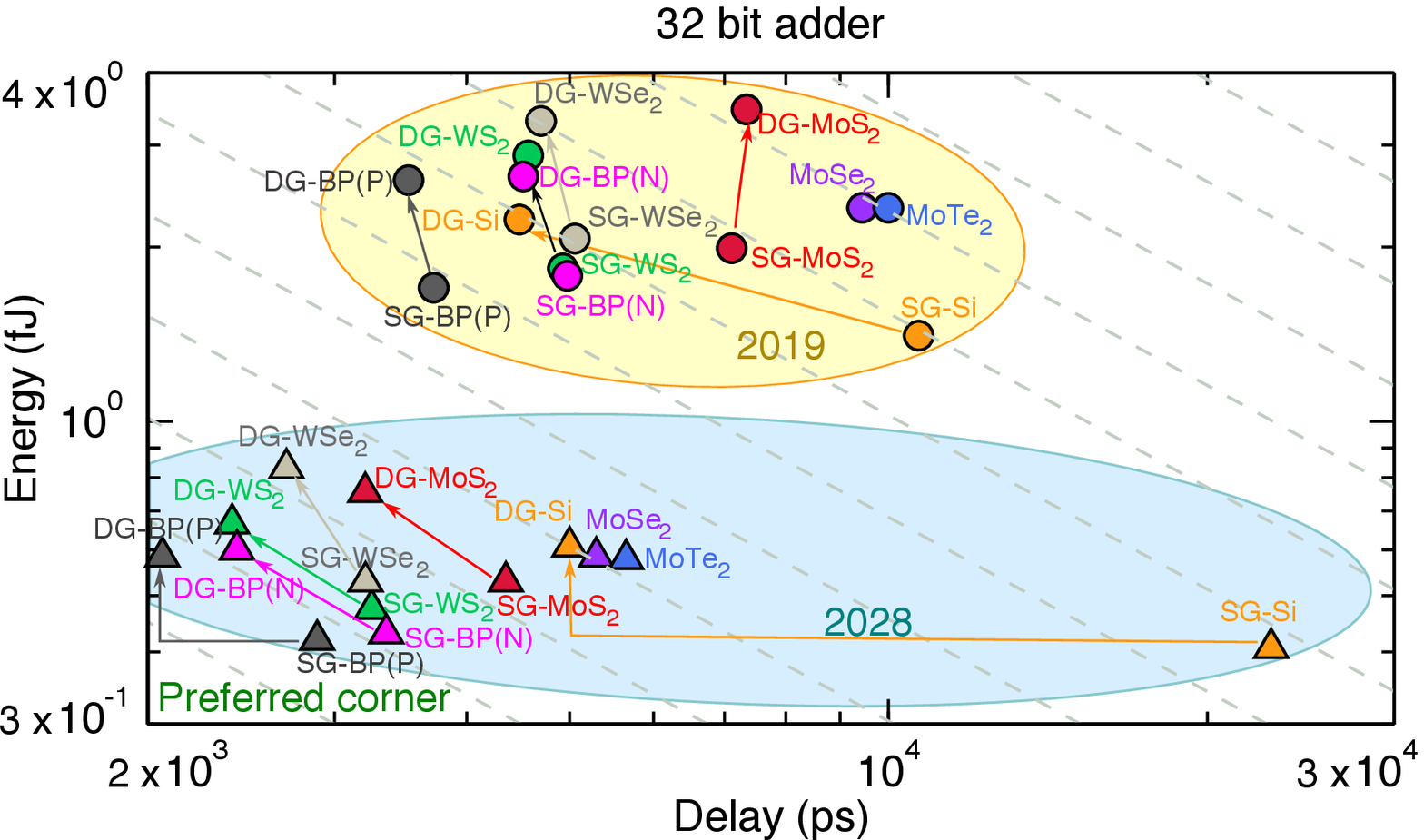}
\caption{Switching energy vs. delay for 32 bit adder. 
\cite{note:bigfig}
}
\label{fig:EvsT}
\end{figure}

The power density as a function of computational throughput is shown in Fig. \ref{fig:thruput}.
Computational throughput 
is defined as number of integer operations per second per unit area 
(32 bit additions in the case of 32 bit adder) \cite{young_nikonov_bcb2}. 
The throughput is the inverse of the circuit delay time in Fig. \ref{fig:EvsT} divided by the 
circuit area.
Since the areas for all adders at a given node are taken to be the same,
the throughput is proportional to the inverse of the adder delay time.
At the 2028 node, SG WSe$_2$, WS$_2$, and BP all have significantly
higher throughputs than DG-Si with slightly higher power density.
Following Refs. \cite{young_nikonov_bcb2} and \cite{young_nikonov_bcb3}, 
we set the power density limit to 10 W/cm$^2$.
All of the FETs lie within the power 
density constraints since they all operate at low voltage (0.3 V).

\begin{figure}
\centering
\includegraphics[width = 3.2in]{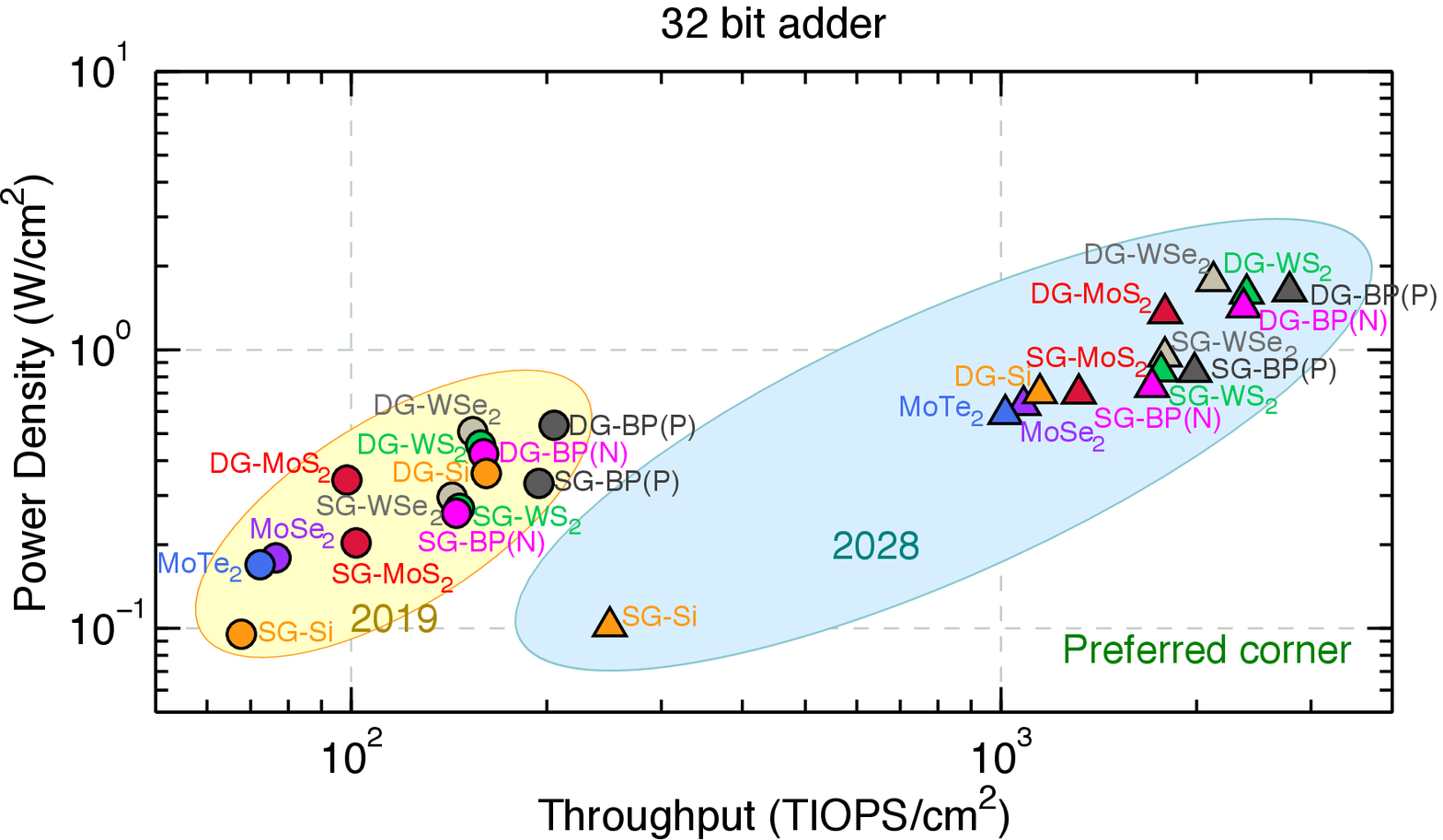}
\caption{Dissipated power vs. computational throughput in tera integar 
operations per sec (TIOPS) per cm$^2$. 
\cite{note:bigfig}
}
\label{fig:thruput}
\end{figure}

\section{Summary and Conclusions}
\label{sec:summary}
We performed quantum mechanical simulations for vdW FETs 
with monolayer MoS$_2$, MoSe$_2$, MoTe$_2$, WS$_2$, WSe$_2$, and BP channels
operating in the LV regime for geometries corresponding to those of the
2019 node and the 2028 node of the 2013 ITRS.
A UTB Si FET was simulated using the same approach to provide a comparison.
The FET serves as a baseline device for determining targets for material parameters.
As the gate length is scaled from 13.3 nm to 5.9 nm, 
blocking the leakage current becomes more critical, 
and the TMD materials with the heavier effective masses benefit 
most from extreme scaling.
For all materials, the ballistic current always reduces with scaling
in agreement with previous work \cite{Klimeck_xsistor_roadmap4_ITRS_2013}.
However, the full current that includes the effect of scattering can either increase or decrease,
and the increase or decrease is governed by two 
competing processes that are both closely tied to the
effective mass, direct tunneling through the channel and backscattering from the channel.
There is an optimum effective-mass of $\sim 0.4$ $m_0$ corresponding to that
of WSe$_2$ that provides a maximum drive current for LV operation
with $V_{DD} = 0.3$ V. 
The short 6 nm gate length combined with 
LV operation is forgiving in its requirements for material quality and contact resistances. 
Low-voltage results in low current and thus low IR drop across the contact resistances,
and the short 6 nm gate length becomes less than the mean free path of the low-mobility
material. 
At the 2028 node, the single gate vdW FETs 
show competitive performance 
in terms of drive current and power density. 
These performance metrics are obtained using currently measured values for mobilities 
shown in Table S1 and
contact resistances shown in Table \ref{tab:Ion} that are comparable to the best measured
contact resistances \cite{MoS2_nmat_phase_engg_contact}.

%
%
%
%
%
%
%
%

\begin{center}
{\Huge Supplementary Material}
\end{center}

{Table} \ref{tab:mat_prop} 
provides the material parameters used in the calculations described in Sec. II of the
paper.
The measured mobility in monolayer MoTe$_2$ was unknown at the time of this work, 
hence  $\mu_{MoTe_{2}}$  was calculated as
$\mu_{MoSe_{2}} \times \frac{(m^*_{MoSe_{2},K})^2}{(m^*_{MoTe_{2},K})^2}$. 
During the review process, we became aware of measurements on multilayer MoTe$_2$ 
flakes showing a room temperature mobility of approximately 21 $\rm{(cm^2/Vs)}$ \cite{ETutuc}.
Since TMDs are weakly coupled van der Waals layers, these mobility values can be
representative of monolayer mobility as well \cite{fallahazad2016PRL} {(or an upper bound).} 
{
One caveat with these values is that the devices were not encapsulated during measurements 
and hence the mobility values represent a lower bound.}
{A mobility value of 21 $\rm{(cm^2/Vs)}$ (in contrast to 42.74 $\rm{(cm^2/Vs)}$ as used in this work) would shorten the mean free path further and degrade the overall performance for MoTe$_2$.}
{
For Si, we followed} Ref. \cite{Venugopal_JAP03} {and used a mobility of 200 $\rm{(cm^2/Vs)}$ 
which could be considered as an optimistic value, 
since at lower inversion charge densities, mobilities can be reduced by a factor 2} \cite{xu2012EDL}. 

\begin{center}
\begin{figure}
\centering
\includegraphics[width = 3.2in]{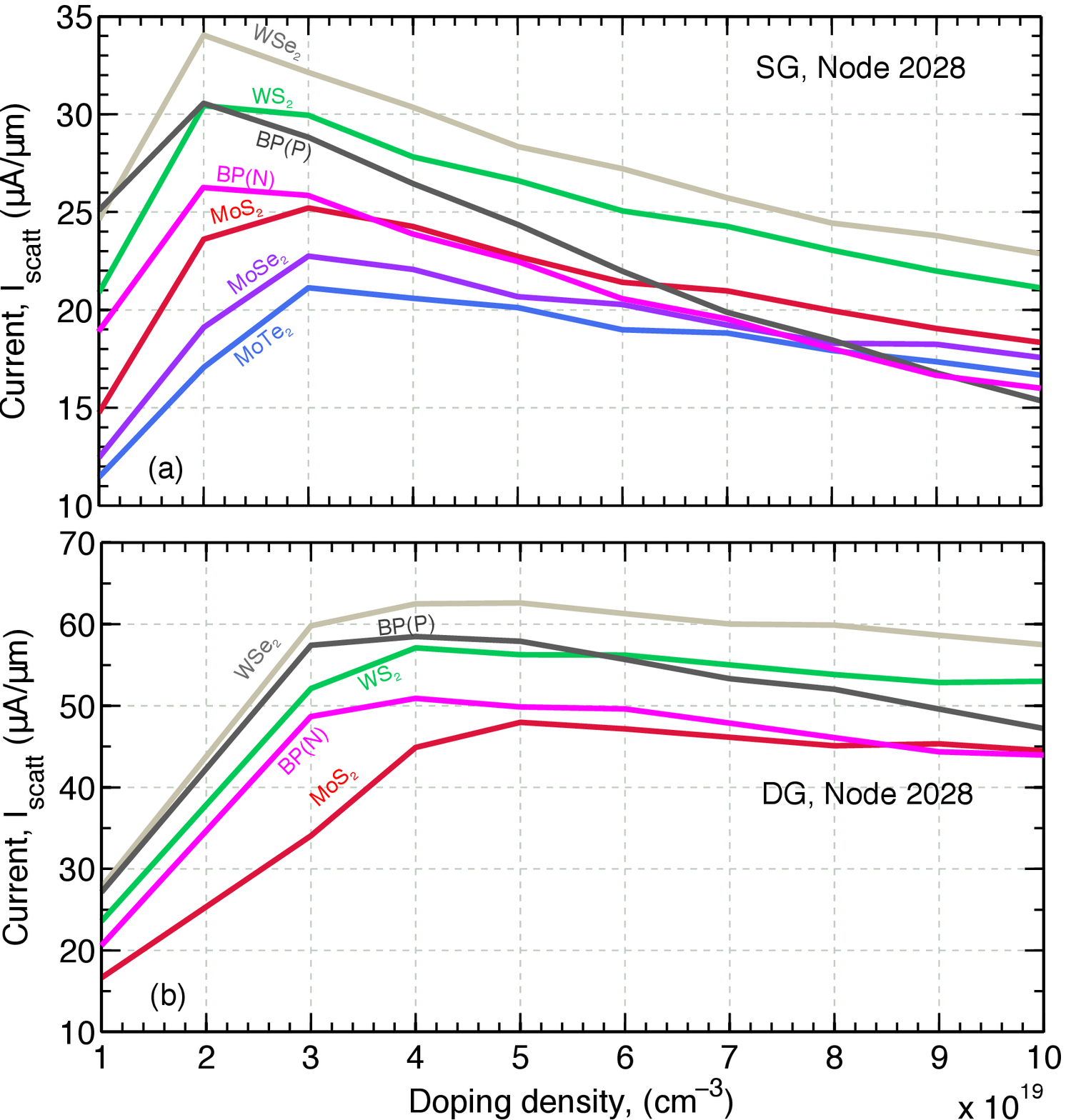}
\caption{On current vs. doping densities for the vdW materials for (a) SG (b) DG configuration at the 2028 node. Each material shows a clear peak at different doping densities which shift toward higher densities for heavy mass materials.}
\label{fig:IvsNd}
\end{figure}
\end{center}

Fig.  \ref{fig:IvsNd} 
shows how the drive currents vary as a function of the source and drain doping densities. 
The doping densities that gave the maximum drive currents (in the absence
of contact resistance) were chosen.
At the highest doping of $1\times10^{20}$ cm$^{-3}$, 
the Fermi level lies close to the band edge for all of the vdW materials.
The source (and drain) degeneracy $E_{Fs}-E_{cs}$ ($E_{vs}-E_{Fs}$) 
varies between {-21.65 meV} to 10.5 meV 
for the vdW materials where $E_{Fs}$ is the source Fermi energy and 
$E_{cs}$ ($E_{vs}$) are the source conduction (valance) band edges.  
Even though the source doping of Si is one order of magnitude less
than the  highest doping used for the TMD FETs, the source degeneracy of the 
Si
Fermi level, $E_{Fs} - E_{cs} \approx 35$ meV, is the largest among all of the FETs.
This is a result of its density of states mass ($0.22~ m_0$) times its degeneracy,
2 orbitals and 2 spins, being the smallest.
For comparison, BP has the smallest transport mass, but, because of its huge anisotropy,
its density of states masses of $0.98~ m_0$ {for hole and $0.44~ m_0$ for electron} are large. 

While source exhaustion sets the lower limit on the doping in an unconstrained
layout, there are design rules that limit the extent of the depletion
regions into the source and drain.
The source and drain depletion lengths will be terminated at the $n^+$ vias
for the metal 1 contacts to the source and drain.
Following the layout of Fig. 26 in Ref. \cite{young_nikonov_bcb2},
for the 2028 node, 
these regions will be 7.1 nm to the left and right of the physical gate
limiting the depletion lengths to 7.1 nm into the source and drain.
To determine whether the layout constraint at the 2028 node
affects the performance trends,
we simulate the SG 
{p-type}
BP and WSe$_2$ FETs 
with $1\times 10^{20}~ \rm cm^{-3}$ doping 
in the via regions on the left and right side of the gate 
with optimized doping between the via and the gate.
The value of the optimum doping does not change, and
the band diagrams for the 
{p-type}
BP FET with and without the heavily doped via
are shown in Fig. \ref{fig:BP_Ec_2e19_1e20_ab}.
The currents for both the WSe$_2$ FET and the BP FET slightly decrease.
For WSe$_2$, $I_{\rm ball} = 59.2$ $\mu$A$/\mu$m, and $I_{\rm scatt} = 32.3$ $\mu$A$/\mu$m.
For 
{p-type}
BP, $I_{\rm ball} = 53.3$ $\mu$A$/\mu$m, and $I_{\rm scatt} = 28.0$ $\mu$A$/\mu$m. 
In both cases $I_{\rm ball}$ decreases more than $I_{\rm scatt}$. 
The reason is that in the on-state, there is a stronger pull
on the channel from the heavily doped drain via that drives the
point at which the channel potential drops by $k_BT$ back
towards the source. 
This reduces the effective channel length, $L_{\rm eff}$,
which reduces the backscattering coefficient $r_c$.
Since the trends and relative performance are not affected by the
proximity of the via we did not consider it in the main text.

For 2019 UTB Si, going to a DG structure increases 
$I_{\rm ball}$ by a factor of 3.9 and $I_{\rm scatt}$ by a factor of 3.7 compared to their
values in the SG geometry.
For 2028 UTB Si,  going to a DG structure increases 
$I_{\rm ball}$ by a factor of 7.3 and $I_{\rm scatt}$ by a factor of 6.7 compared to their
values in the SG geometry.
The much larger increases in the UTB Si currents going from a SG to a DG geometry
at the 2028 node compared to those of the 2D material currents are a result of the different
channel thicknesses. 
At the 2028 node, a double gate is required to control the potential through
the 3 nm Si channel. 
This is illustrated in Fig. \ref{fig:Si_Ec}.
The set of green curves in Fig. \ref{fig:Si_Ec} shows the conduction band edges
for SG Si at each grid point through the depth of the Si channel. 
The highest curves are at the top of the channel adjacent to the gate oxide, and
the lowest curves are at the bottom of the channel adjacent to the substrate.
The large spread in energy of the curves illustrates the loss of control of the 
channel potential by the single gate.
The set of blue curves show the same set of conduction band edges for the DG
device. 
The double gate provides good control of the potential throughout the channel.
For the thinner monolayer vdW FETs, a single gate is adequate.
Figs.  \ref{fig:devEvsT} - \ref{fig:thruput} 
are enlarged versions of Figs. 7 - 9 of the main article.

\begin{center}
{\bf NEGF Details}
\end{center}

The heavily doped source and drain regions are treated as contacts in equilibrium with
their respective Fermi levels \cite{Klimeck_APL95}, and the charge in those regions is calculated from
the equilibrium expression,
\begin{equation}
n_i(p_i) = \sum_{\nu} s_\nu n_\nu \sqrt{\frac{m_y^\nu k_B T}{2 \pi \hbar^2}}\int \frac{dE}{2\pi} [A_i^\nu F_{-1/2}(\eta_{S(D)})] 
\end{equation}
where $\nu$ is the band index, $s_\nu$ is the spin degeneracy and $n_\nu$ is the valley degeneracy which is 2 for the $K$-valleys, 
6 for the $\Lambda$-valleys and 1 for the $\Gamma$-valley. 
$m_y^\nu$ is the effective mass in the width direction, 
$k_B$ is Boltzmann's constant, $T$ is temperature, and $A_i^\nu$ 
is the spectral function 
on site $i$ for band $\nu$ given by 
$-2 {\rm Im} G^R_{i,i;\nu} (E)$.
The factors $\eta_{S(D)}=(\mu_{S(D)}-E)/k_BT$ are the reduced Fermi factors
resulting from analytically integrating over the transverse momentum
where $\mu_{S(D)}$ is the Fermi level of the source (drain), respectively.
Within the device region, the charge is calculated from the 
non-equilibrium expression,
\begin{align}
n_i(p_i) =&\sum_\nu s_\nu n_\nu \sqrt{\frac{m_y^\nu k_B T}{2 \pi \hbar^2}} \; \cdot
\nonumber \\
& \int \frac{dE}{2\pi} [A_{i;S}^\nu F_{-1/2}(\eta_S) + A_{i;D}^\nu F_{-1/2}(\eta_D)],
\label{eq:n_neq}
\end{align}
where $A_{i;S(D)}^\nu$ is the source (drain) connected spectral function at site $i$ for band $\nu$, 
given by
$A_{i;S}^\nu = | G^R_{i,1;\nu} |^2 \Gamma_{1,1;\nu}$ and
$A_{i;S}^\nu = | G^R_{i,N;\nu} |^2 \Gamma_{N,N;\nu}$.

The drain current is calculated within the self-consistent loop from
\begin{align}
I_D =&\sum_\nu \left(\frac{1-r_c}{1+r_c}\right) s_\nu n_\nu \left(\frac{q}{h}\right) \sqrt{\frac{m_y^\nu k_B T}{2 \pi \hbar^2}} \; \cdot
\nonumber \\
& \int \frac{dE}{2\pi} T_{\nu}(E) [F_{-1/2}(\eta_S) - F_{-1/2}(\eta_D)],
\label{eq:id}
\end{align}
where $T_{\nu}(E)$ is the transmission coefficient for band $\nu$, 
and $r_c$ is the backscattering coefficient, 
$r_c = L_{\rm eff}/(L_{\rm eff}+\lambda)$ \cite{lundstrom_elem_backscattering_th}.
$L_{\rm eff}$ is the critical length determined by the distance from top of the barrier in the channel to the position towards the drain where the potential drops by $k_BT/q$.
The mean free path $\lambda$ is calculated using \cite{lundstrom_backscattering}
\begin{equation}
\lambda = \frac{2k_BT}{q}\frac{\mu_0}{v_T} \frac{F_0(\eta_{top})}{F_{-1/2}(\eta_{top})}
\end{equation}
where $v_T$ is the thermal velocity and $\eta_{top} = (\mu_S - E_{c,top})/k_BT$.

The same equations are used for the UTB Si FET. 
The discretization within the 3 nm
thick channel is 0.2 nm in the $z$ direction and 0.25 nm in the $x$
direction as shown in Fig. 1 of the main article.
The charge is calculated everywhere using the non-equilibrium
expression, Eq. (\ref{eq:n_neq}).
%
%

\beginsupplement
\begin{table*}
\caption{
Properties for all the materials studied in this work 
\cite{Falko_allTMDC_bandstructure, MoS2_1L_natcomm, MoSe2_monolayer_acsnano, WS2_1L_BNsandwich, WSe2_DJena_Nano_Letters, BP_1L_sdas_nanolett, kalam2012mos2,Venugopal_JAP03, BP_1L_extracted_mu_nat_comm}. 
All materials except BP are n-type.
$\mu_0$ is the measured low field mobility.
CB and VB stand for conduction and valance band respectively. The $\Lambda$-valley lies roughly half-way between the $\Gamma$ and $K$  valleys. 
$\Delta E_{K\Lambda}$ is the energy difference between the $K$-valley and the $\Lambda$ valley. 
Subscript $s-s, e$ and $h$ stand for spin-split, electron and hole respectively. 
Effective masses along different directions in anisotropic valleys are represented by appropriate subscripts.  
}
\label{tab:mat_prop}
\vspace*{-18pt}
\begin{center}       
\vspace*{8pt}
\begin{tabular}{|c|c|c|c|c|c|c|c|} 
\hline
\tr Material                             &MoS$_2$        &MoSe$_2$  &MoTe$_2$ &WS$_2$   &WSe$_2$  &BP     &Si\\
\hline
\tr Bandgap (eV)                     & 1.67              &1.40           &0.997        & 1.60       &1.30         &1.55 &1.12\\
\hline
\tr Thickness (nm)                  & 0.65              & 0.73          & 0.7           & 0.62       & 0.73        & 0.57          & 3\\
\hline
\tr Measured mobility,               & 81                & 50              & 42.74$^{\dagger}$         
                                                                                                                 & 185      &202         & $\mu_e$ = 94   & 200\\
\tr $\mu_0 \rm{(cm^2/V-s)}$ &&&&&& $\mu_h$ = 116 & \\
\hline
\tr Effective                & \mk =0.43   & \mk =0.49 & \mk =0.53 & \mk =0.35   
                                                                                                                                &\mk =0.39
                                                                                                                                                 &$m^*_{e,\Gamma,x}=0.17$                                                                                                                     & $m^*_l = 0.98$\\
\tr mass, $m^*(\times m_o)$  & \mkss =0.46 & \mkss =0.56 
                                                                                            & \mkss =0.62 
                                                                                                                 & \mkss =0.26 
                                                                                                                                &\mkss =0.28
                                                                                                                                                & $m^*_{e,\Gamma,y}=1.12$                                                                                                                     & $m^*_t = 0.22$\\
\tr                                           & \mSx =0.56   & \mSx =0.48   
                                                                                           & \mSx =0.43& \mSx =0.52 
                                                                                                                                &\mSx =0.42& $m^*_{h,\Gamma,x}=0.15$ & \\
\tr                                           & \mSy =1.13   &\mSy=1.08 & \mSy =0.99& \mSy =0.74 
                                                                                                                                &\mSy =0.74& $m^*_{h,\Gamma,y}=6.35$ & \\
\tr                                           & \mSssx =0.64 & \mSssx =0.54 
                                                                                            & \mSssx =0.42 
                                                                                                                  & \mSssx =0.69
                                                                                                                                &\mSssx =0.73&&\\
\tr                                           & \mSssy =1.21 & \mSssy =1.11 
                                                                                           & \mSssy =1.16
                                                                                                                  & \mSssy =0.94 
                                                                                                                               &\mSssy =0.91&&\\
\hline
\tr CB $K$-valley                    & 3                     & 22           & 36                & 32      &37                  & $\times$       & $\times$\\
\tr spin-splitting (meV)           &                       &                 &                     &           &                      &&\\
\hline
\tr CB $\Lambda$-valley           & 70                   & 21           & 22                & 264    &218                & $\times$ & $\times$\\
\tr spin-splitting (meV)           &                       &                  &                    &           &                      &&\\
\hline
\tr CB $\Delta E_{K\Lambda}$ (meV) & 207        & 137           & 158              & 81      &35                  & $\times$ & $\times$\\
\hline
\tr Mean free path,             &\ldaK= 5.2     
                                                                        &\ldaK= 3.4   
                                                                                            &\ldaK= 3.1     
                                                                                                                   &\ldaK= 10.7           
                                                                                                                                &\ldaK= 12.3
                                                                                                                                                        & $\lambda_{CB}$ = 11.1   &  9\\
$\lambda$ (nm) (approx.)       &\ldaKss= 5.3                    
                                                                       &\ldaKss= 3.6                    
                                                                                            &\ldaKss= 3.3
                                                                                                                   &\ldaKss= 9.1    
                                                                                                                                 &\ldaKss= 10.3 
                                                                                                                                                       & $\lambda_{VB}$ = 6.1  &\\
\tr                                          &\ldaS= 7        &\ldaS= 4.1
                                                                                           &\ldaS= 3.3     &\ldaS= 14
                                                                                                                                 &\ldaS= 14.5
                                                                                                                                                      &&\\
\tr                                          &\ldaSss= 7.3 
                                                                      &\ldaSss= 4.2
                                                                                          &\ldaSss= 3.4
                                                                                                                   &\ldaSss= 16
                                                                                                                               &\ldaSss= 17.5
                                                                                                                                                    &&\\
\hline
\multicolumn{8}{l}{\footnotesize{$^{\dagger}$Measured mobility in monolayer MoTe$_2$ was unknown, hence  $\mu_{MoTe_{2}}$  is calculated as: $\mu_{MoSe_{2}} \times \frac{(m^*_{MoSe_{2},K})^2}{(m^*_{MoTe_{2},K})^2} $} 
}\\
\end{tabular}

\end{center}
\end{table*}

\clearpage

\begin{center}
\begin{figure}
\centering
\includegraphics[width = 4.0in]{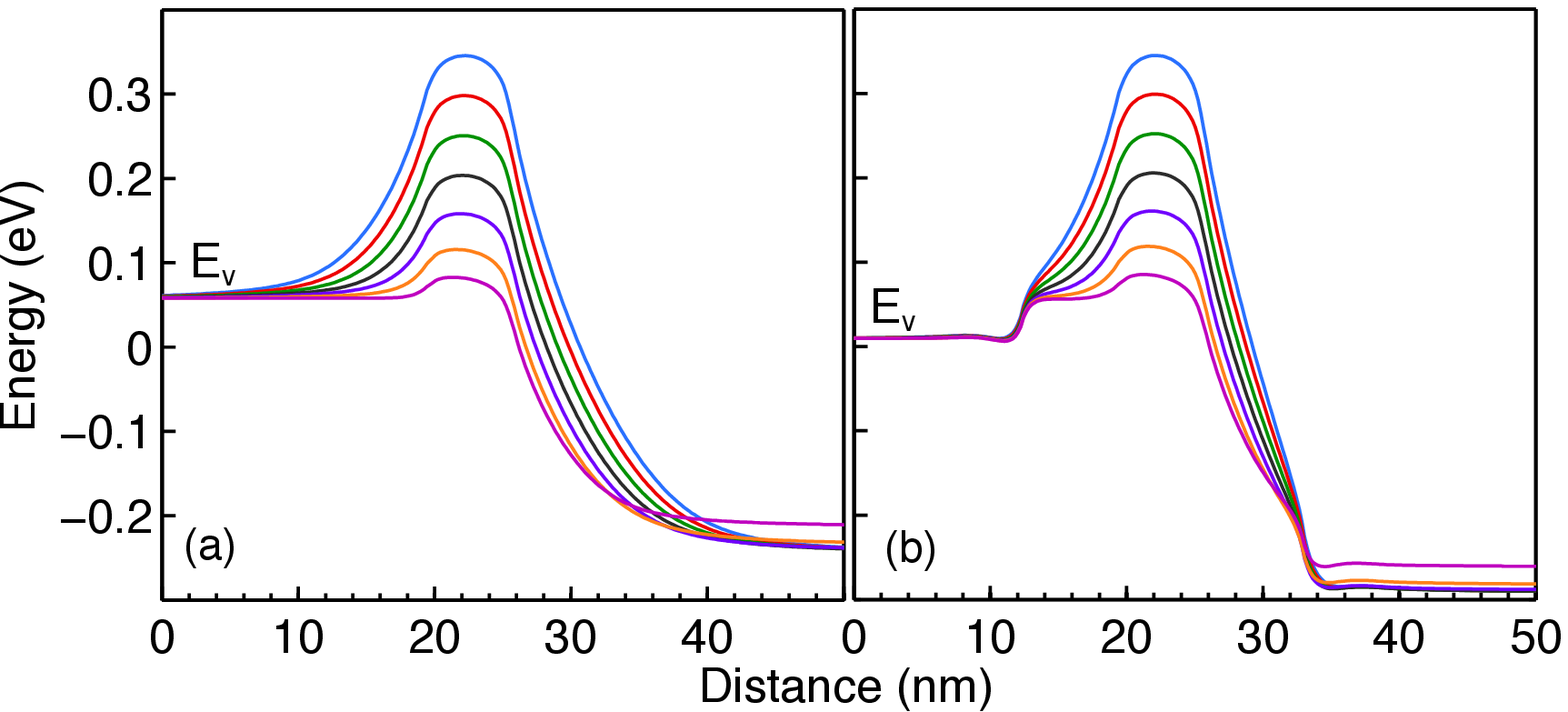}
\caption{
Energy band diagrams for different gate biases of the 
{p-type}
BP FET at the 2028 node.
{Kinetic energy for the holes is taken to be positive.}
The source Fermi energy is 0 eV.
(a) $2\times 10^{19}$ cm$^{-3}$ doping throughout the source and drain regions.
(b) $2\times 10^{19}$ cm$^{-3}$ doping in the source and drain regions
with $1\times10^{20}$ cm$^{-3}$ doping starting at the edge of the source and drain vias.
}
\label{fig:BP_Ec_2e19_1e20_ab}
\end{figure}

\begin{figure}
\centering
\includegraphics[width = 4.0in]{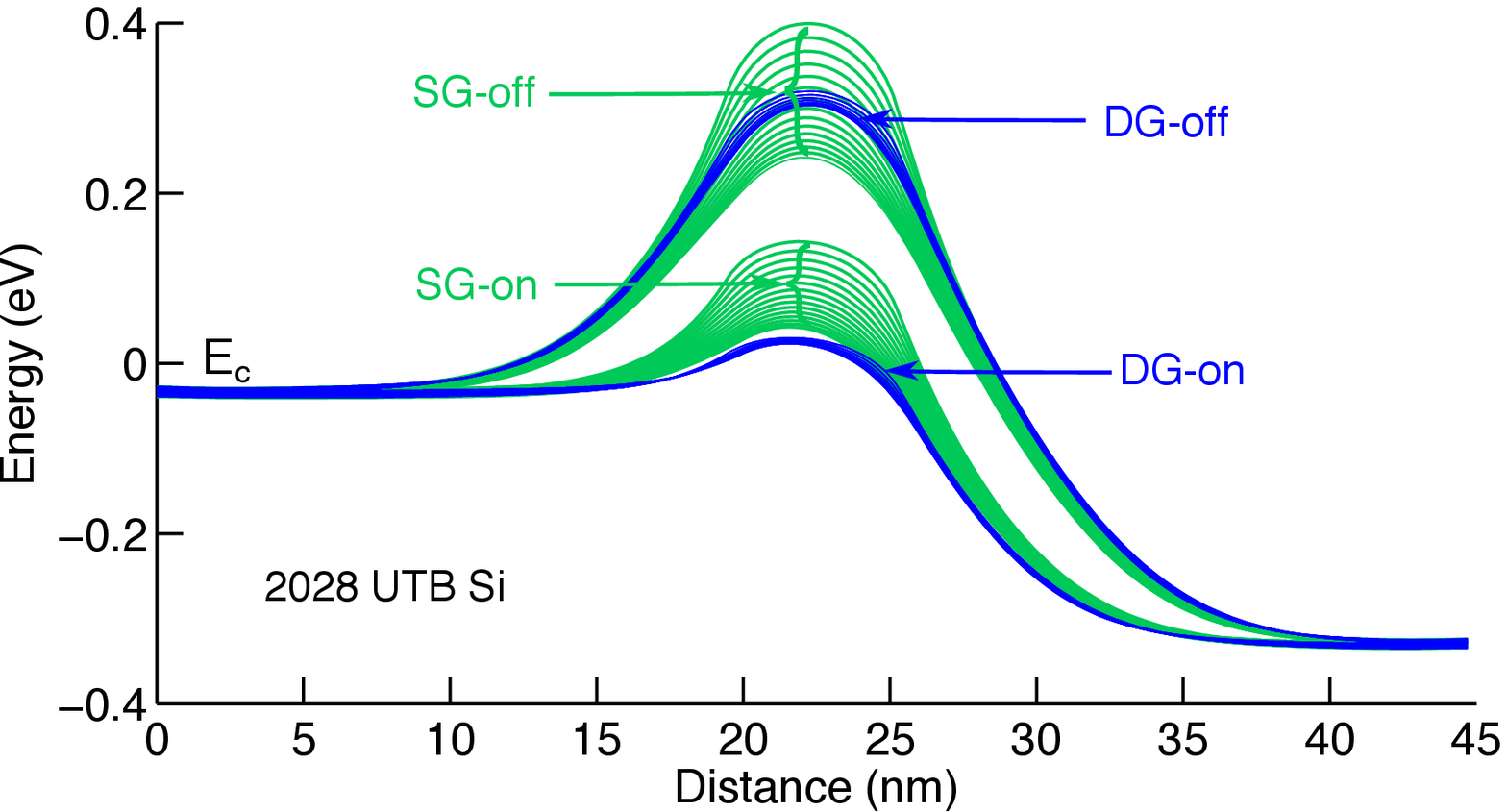}
\caption{(Color online)
2028 UTB SG and DG Si conduction band edges in the on and off states. 
The source Fermi energy is at 0 eV.
The green curves are the conduction band edges taken at each grid point in the SG Si
channel. The highest curve is closest to the gate, and the lowest curve
is closest to the substrate.
The blue curves are the same for the DG structure.
The lack of spread shows that the DG provides good control of potential
throughout the channel. 
}
\label{fig:Si_Ec}
\end{figure}
\end{center}
\clearpage
%
%
%
%
\begin{figure*}
\centering
\includegraphics[width = 6in]{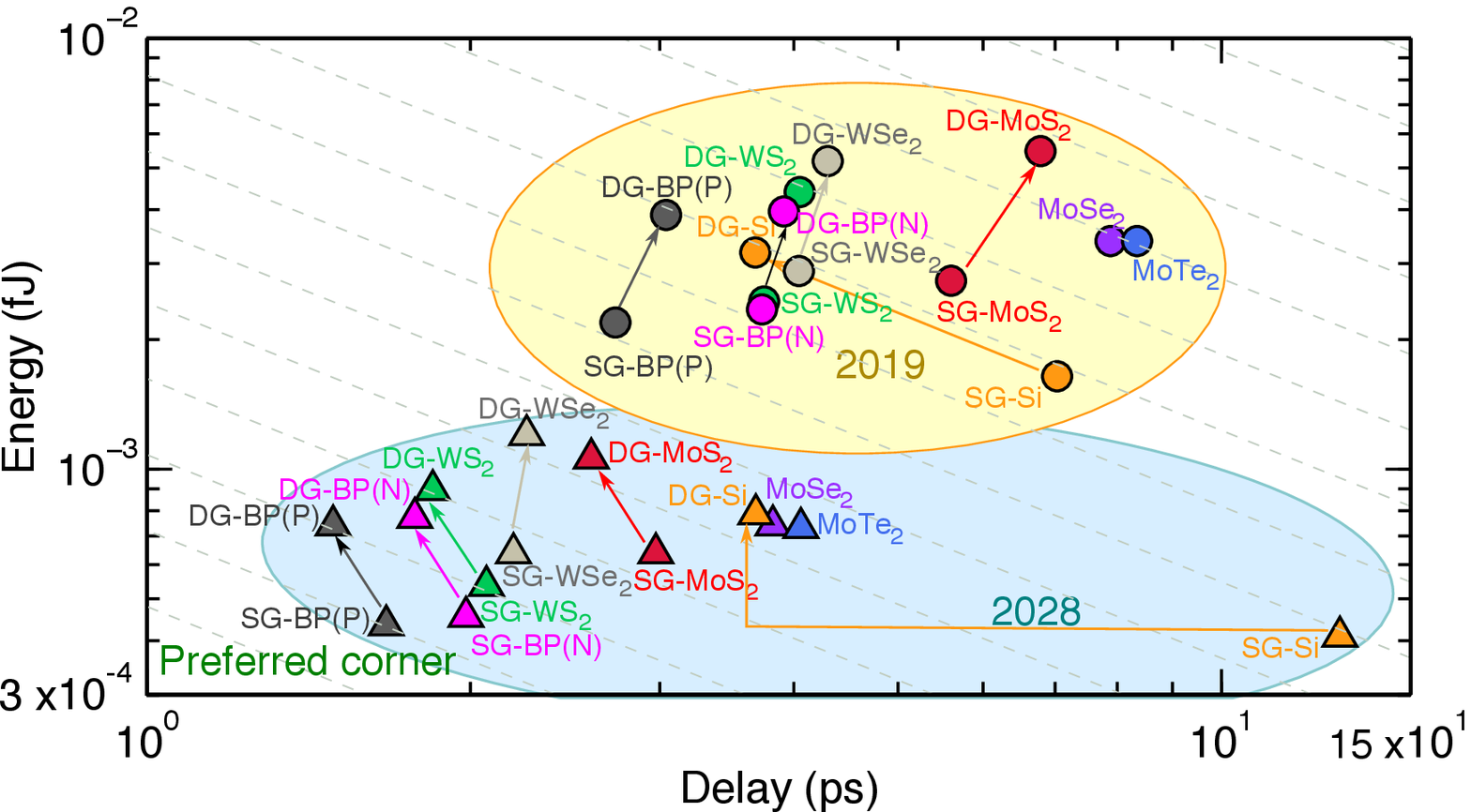}
\caption{Intrinsic switching energy versus delay for individual FETs. 
Circles and triangles stand for 2019  and 2028 node, respectively.
Diagonal dashed lines are constant energy-delay product lines.
Each successive line represents an increase of 1.5.
}
\label{fig:devEvsT}
\end{figure*} 

\begin{figure*}
\centering
\includegraphics[width = 6in]{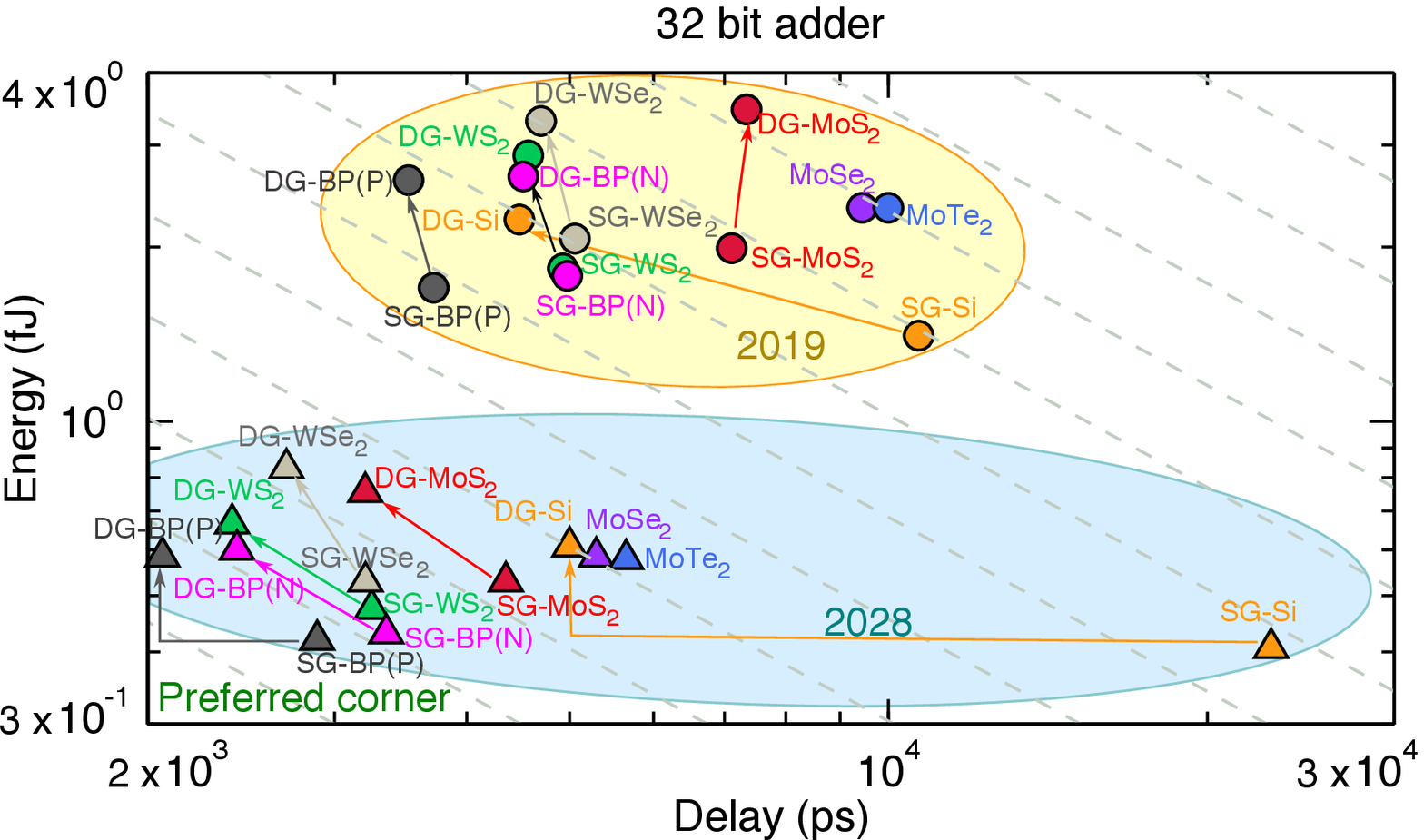}
\caption{Switching energy versus delay for 32 bit adder.}
\label{fig:EvsT}
\end{figure*}

\begin{figure*}
\centering
\includegraphics[width = 6in]{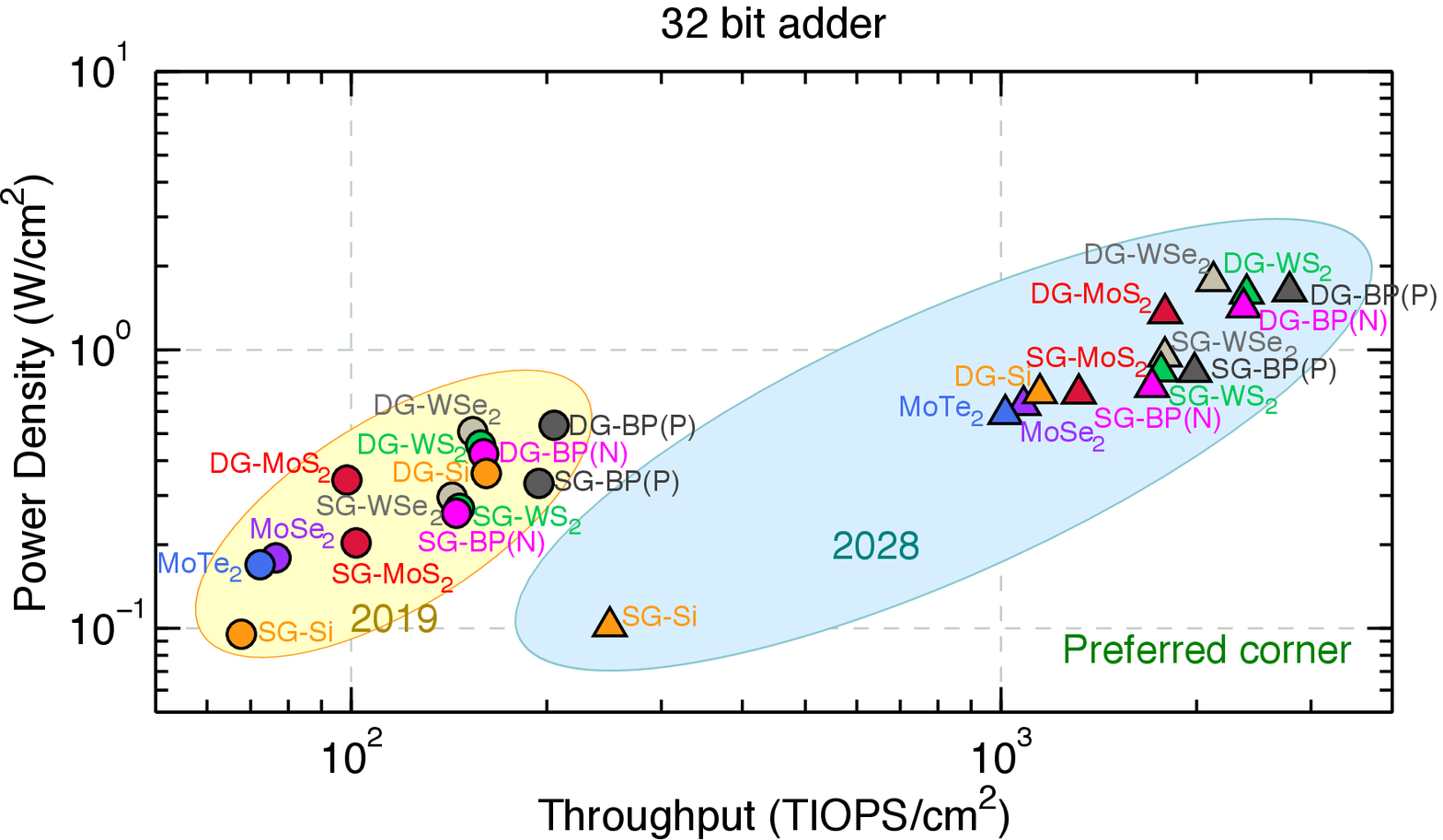}
\caption{Dissipated power vs. computational throughput in tera integar operations per sec 
(TIOPS) per cm$^2$.}
\label{fig:thruput}
\end{figure*}

\clearpage

%

%
\bibliography{sylvia}
\bibliographystyle{IEEEtran}



\end{document}